\documentclass[aps,prd,onecolumn,amsmath,11pt,superscriptaddress,floatfix,nofootinbib,preprintnumbers]{revtex4-2}

\usepackage{mathrsfs}
\usepackage{amsfonts}
\usepackage{amsmath,amssymb,bm}
\usepackage{array}
\usepackage{verbatim}
\usepackage{epsfig}
\usepackage{graphicx}
\usepackage{hyperref}
\hypersetup{colorlinks, linkcolor = [rgb]{0, 0, 0.5}, citecolor = [rgb]{0,0.0,0.5}, urlcolor = [rgb]{0,0.0,0.5}}
\usepackage[normalem]{ulem}
\usepackage{xcolor}
\usepackage{ulem}
\usepackage{multirow}

\usepackage{subcaption}
\usepackage{graphicx}
\graphicspath{{./figs/}}
\usepackage{dcolumn}
\usepackage{bm}
\usepackage{slashed}
\usepackage{siunitx}
\usepackage{ulem,xpatch}
\usepackage{hyperref}
\usepackage[mathlines]{lineno}
\usepackage{comment}
\usepackage{soul}
\usepackage{xcolor}
\usepackage{xfrac}

\allowdisplaybreaks[4]

\begin{document}

\title{Direct Access to Nuclear Photon TMDs with In Situ Probes in Ultraperipheral Heavy-Ion Collisions}

\author{Cong Li}
\affiliation{School of Information Engineering, Zhejiang Ocean University, Zhoushan, Zhejiang, China}
%


\begin{abstract}
Resolving local structure in photon distributions is an important goal in probing strong coherent electromagnetic fields. We propose using electron--positron pairs produced in situ in ultraperipheral heavy-ion collisions as dynamical probes to directly extract photon distributions. We define a theoretical reference cross section using the complete tree-level amplitude for two-photon-induced radiative lepton-pair production. When background effects factorize into a scalar Bose-enhancement factor, the ratio of the background-modified to reference differential cross sections, minus unity, yields the photon occupation number at a fixed momentum. Finite bins yield averages weighted by the reference cross section. We specify the conditions under which kinematic selections, acceptance, and radiative corrections preserve this local inversion.
\end{abstract}

\maketitle

\section{Introduction}

Ultraperipheral heavy-ion collisions provide both strong coherent electromagnetic fields and charged leptons produced through two-photon reactions \cite{Bertulani_2005}. Existing measurements of lepton-pair momentum and angular distributions demonstrate that these final states carry information about the structure of nuclear photon fields~\cite{PhysRevLett.127.052302,Aad2023}. Bose-statistical enhancement has been observed in heavy-ion collisions through identical-pion correlations \cite{PhysRevC.71.044906,PhysLettB.696.328}. Occupation-dependent stimulated gluon emission has also been studied in calculations of parton energy loss in a quark–gluon plasma \cite{PhysRevLett.87.142301}. Here we investigate how final-state Bose enhancement in radiative observables can be used to extract the background photon occupation number at fixed momentum and subsequently relate it to nuclear photon distributions.

We propose using Breit--Wheeler lepton pairs produced in the collision as dynamically generated, in situ charged probes of the momentum structure of the surrounding nuclear photon field. Specifically, starting from the extensively studied two-photon process $\gamma\gamma\to e^+e^-$ \cite{Adams2004}, we use the produced leptons as probes and consider the additional real photon in the radiative channel $\gamma\gamma\to e^+e^-\gamma$. This photon carries information acquired as the leptons propagate and radiate in the background electromagnetic field. Within the factorized description developed here, the background photon occupation number enters the cross section multiplicatively and can be separated from a process-dependent, precisely calculable QED hard reference. The conventional forward mapping can therefore be converted into a local inversion,
\begin{equation}
d\sigma
\;\longrightarrow\;
n_\gamma(y_\gamma,k_\perp)
\;\longrightarrow\;
f_1^{\gamma/A}(x,k_\perp).
\end{equation}
With consistent matching of photon states and normalization conventions, the extracted occupation number can further be related to the nuclear photon distribution.

The central theoretical challenge is that the probes are generated dynamically by the collision itself rather than supplied by an external beam. The theoretical hard reference must therefore consistently include the nuclear photon luminosity and the complete $\gamma\gamma\to e^+e^-\gamma$ amplitude, including electron and positron radiation and their interference, while excluding the background occupation factor to be extracted. Within our framework, comparison with this complete-process reference separates spectral structures arising from probe production and radiation dynamics from the additional enhancement associated with background occupation.

The extraction requires no assumed global parametrization of the target distribution, allowing occupations at different photon momenta to be tested independently. In principle, this direct approach can therefore preserve local structures, non-Gaussian behavior, or other fine transverse-momentum features that may be averaged out in conventional convolution-based analyses. Such features can signal additional physical mechanisms or effects. The method can both identify local structures and constrain structures of specified scales and amplitudes that remain unobserved within the available resolution and measurement precision. Its value thus does not depend on assuming that the underlying distribution contains fine structure.

We also examine how kinematic selections and radiative corrections affect the extraction, characterize the weighted averaging induced by finite bins and detector momentum migration, and propose consistency tests at the same photon momentum using different lepton selections. This work focuses on establishing the extraction method, its applicability conditions, and its testable predictions. The radiative lepton-pair process used as a concrete implementation merits further systematic study of its radiation dynamics, kinematic optimization, and experimental prospects. The in situ probe approach developed here in QED also suggests a possible route toward probing fine structure in nuclear color fields with quark pairs \cite{PhysRevD.87.054010}. The greater complexity of QCD dynamics, however, presents additional theoretical and experimental challenges \cite{PhysRevLett.106.132001}.

\section{In Situ Lepton Probes and the Background Model}
\label{2}

We consider coherent radiative lepton-pair production in ultraperipheral collisions of two nuclei, both of which remain intact. The nuclear electromagnetic fields supply two initial quasireal photons. The Breit--Wheeler channel dynamically generates charged lepton pairs that serve as in situ probes in the corresponding radiative channel,
\begin{equation}
\gamma(k_1)+\gamma(k_2)
\to
e^-(p_1)+e^+(p_2)+\gamma(k).
\end{equation}
Probe production and radiation are described jointly by the complete three-body amplitude. Our primary observable is the final-state photon momentum, specified by $(y_\gamma,k_\perp)$, where $k_\perp=|\boldsymbol{k}_\perp|$ and $E_\gamma=k_\perp\cosh y_\gamma$. All remaining unobserved initial- and final-state degrees of freedom are collectively denoted by $\Phi$. We select the correlation region with small-transverse-momentum photons close to the beam axis and central leptons. The detailed kinematics and the rationale for this selection are given in Appendix~B.

We do not measure the final-state photon polarization and integrate over the unobserved angular variables within the selected region, with full azimuthal coverage for the photon and leptons. In this leading-power treatment, the kinematic selections, geometric weights, and corrected acceptance introduce no additional lepton-azimuthal weights. As shown in Appendix~B, after summing over the final-state photon polarization and performing the full azimuthal integrations, the initial-state linear-polarization terms do not contribute, leaving only the unpolarized hard function.

The initial nuclear photon distribution, \(x f_1^{\gamma/A}(x,l_\perp)\)—where \(l_\perp\) denotes the transverse momentum of the incident photon—is provided within the framework of the coherent Equivalent Photon Approximation (EPA) \cite{BERTULANI1988299,PhysRevC.47.2308,ZHA2020135089,Klein_2017,Li_2019}. Background occupation numbers are introduced separately via statistical weights associated with photon emission, while the corresponding occupation number are not included in the theoretical reference spectrum. When constructing this reference spectrum, model uncertainties inherent in the EPA input propagate through to the final extracted occupation numbers. These uncertainties are expected to be small: the EPA is widely applied in calculations for ultraperipheral collisions (UPCs), and, when its photon distribution is convolved with hard processes, it successfully describes a variety of measured cross sections and kinematic distributions \cite{Aad2023,PhysRevC.104.024906}. However, agreement at the level of the convolution does not equate to a point-by-point verification of the fine-grained local structure within the underlying photon distribution.

The extraction targets the mean background occupation number of the selected final-state photon mode, defined by the momentum-diagonal element of the background one-photon correlation function. For the selected modes, we adopt an axially symmetric, polarization-independent scalar statistical description and denote the occupation number by $n_\gamma(y_\gamma,k_\perp)$. This description requires neither thermal equilibrium nor an assumed global momentum dependence. For either physical polarization, we define $n_\gamma(k)\equiv\operatorname{Tr}\!\left[\rho_{\mathrm{bg}}\,a_{\lambda,k}^{\dagger}a_{\lambda,k}\right]$. Here, $\rho_{\mathrm{bg}}$ is the normalized background density operator, and $a_{\lambda,k}$ is the normalized mode annihilation operator. Thus, $n_\gamma$ is the dimensionless mean occupation number per physical polarization mode. We define this occupation number in momentum space without explicitly resolving the spatial distribution of the background photon source.

To relate the extracted occupation to a nuclear photon distribution, let $f_\gamma(x,k_\perp)$ denote the photon number distribution with respect to $dx\,d^2\boldsymbol{k}_\perp$. Provided that the background occupation and the nuclear photon distribution refer to the same photon modes and use consistent photon-state, phase-space, and polarization conventions, the mode-counting relation in Appendix A gives
\begin{equation}
f_\gamma(x,k_\perp)
=
V_{\mathrm{match}}\,
\frac{2E_\gamma}{(2\pi)^3x}\,
n_\gamma(k).
\label{match}
\end{equation}
We use on-shell photon kinematics and $x=k^+/\sqrt{s_{NN}}$. The factor $E_\gamma/x$ is the longitudinal phase-space Jacobian, and the “2” accounts for the two physical polarizations. Under the mode matching stated above, $V_{\mathrm{match}}$ is the common normalization coefficient associated with the mode counting in Appendix \ref{app1} and has dimensions of volume. It can be calibrated using absolute two-photon cross sections through Eq. \eqref{matching}, or an EPA reference, to reconstruct \(f_\gamma\) from \(n_\gamma\). This normalization matching is separate from the extraction of \(n_\gamma\) through the cross section ratio.

Motivated by the correlator description of initial-state nuclear photons, we describe tagged-photon emission through the background two-point function. Radiation adds one photon to a background photon mode. The bosonic commutation relation gives $\operatorname{Tr}\!\left[\rho_{\mathrm{bg}}\,a_{\lambda,k}a_{\lambda,k}^{\dagger}\right]=1+n_\gamma(k).$ The unit term represents spontaneous emission, while the occupation term represents stimulated emission. This mode identity does not require spatial uniformity. We adopt a momentum-diagonal scalar approximation with the vacuum on-shell spectral density, assuming that the statistical weight depends only on the tagged-photon momentum and factorizes from the complete hard process. The positive-energy cut photon propagator is then
\begin{equation}
    \begin{aligned}
D_{\mu\nu}^{\mathrm{bg,cut}}(k)
&\equiv
\theta(k^0)
\int d^4r\,e^{ik\cdot r}
\operatorname{Tr}\!\left[
\rho_{\mathrm{bg}}A_\mu(r)A_\nu(0)
\right]
\\
&=
2\pi\,\theta(k^0)\,\delta(k^2)
\left[1+n_\gamma(k)\right](-g_{\mu\nu})
\end{aligned}
\label{bg_pro}
\end{equation}
where $-g_{\mu\nu}$ represents the physical polarization sum. The occupation \(n_\gamma(k)\) is related to the nuclear photon distribution under the mode-matching condition stated around Eq. \eqref{match}.

\section{Complete Hard Reference and Pointwise Inversion}
\label{3}

The theoretical input reference describes $\gamma\gamma\to e^-e^+\gamma$ with the tagged-photon emission evaluated in vacuum, so that its statistical factor is unity. The complete tree-level cross section is convolved with the incident nuclear photon distributions and evaluated with the same event selections. We write the unpolarized squared tree-level amplitude as
\begin{equation}
\overline{|\mathcal{M}_{\mathrm{ref}}|^2}
=
(4\pi\alpha)^3 H_{\mathrm{BW}}.
\end{equation}
For on-shell kinematics with negligible electron mass, the hard function has the compact form \cite{BERENDS198122}
\begin{equation}
H_{\mathrm{BW}}
=
2(p_1\cdot p_2)
\sum_{i=1}^{3}
\frac{
(p_1\cdot q_i)^2+(p_2\cdot q_i)^2
}{
\displaystyle
\prod_{\substack{j=1\\ j\ne i}}^{3}
(p_1\cdot q_j)(p_2\cdot q_j)
}.
\end{equation}
Here, $(q_1,q_2,q_3)=(k_1,k_2,k)$. In the selected axial correlation region, the reference spectrum uses the leading-power limit of this expression, evaluated with photon momenta along the beam axis. The angular integrations follow the observable definition in Sec.~\ref{2}. The kinematics and tensor contractions are detailed in Appendix~\ref{app2}. We impose photon-energy and central-lepton transverse-momentum thresholds, together with photon–lepton angular isolation, to exclude the soft and collinear singular regions of the massless tree-level reference \cite{DITTMAIER200069}.

Using the unpolarized nuclear photon distributions retained after the lepton-azimuthal integrations in Sec.~\ref{2}, the differential reference spectrum is
\begin{equation}
\begin{aligned}
\frac{d\sigma_{\mathrm{ref}}}
{dy_\gamma\,d^2\boldsymbol{k}_\perp}
={}&
\int d\Phi\,
\frac{\alpha^3}{4\pi^2}
\frac{H_{\mathrm{BW}}}{\hat{s}^{\,2}}
\left[
x_1 f_1^{\gamma/A}
(x_1,k_{1\perp}^2)
\right]
\left[
x_2 f_1^{\gamma/B}
(x_2,k_{2\perp}^2)
\right]
\\
&\times
\delta^{(2)}
\!\left(
\boldsymbol{q}_\perp
-\boldsymbol{k}_{1\perp}
-\boldsymbol{k}_{2\perp}
\right).
\end{aligned}
\label{ref_diff}
\end{equation}
where $\boldsymbol{q}_\perp
=\boldsymbol{p}_{1\perp}
+\boldsymbol{p}_{2\perp}
+\boldsymbol{k}_\perp$
and
$d\Phi
=d^2\boldsymbol{k}_{1\perp}\,
d^2\boldsymbol{k}_{2\perp}\,
dy_1\,d^2\boldsymbol{p}_{1\perp}\,
dy_2\,d^2\boldsymbol{p}_{2\perp}$. Longitudinal momentum conservation fixes $x_1$ and $x_2$, and $\hat{s}=x_1x_2s_{NN}$ in the leading-power approximation used for the hard process. Here, the phase-space integration smooths out potential local fluctuations in the incident distributions $f_{1}^{\gamma/A,B}$, so that $d\sigma_{\rm ref}$ depends only on their well-validated global convolution properties.

Under the statistical factorization specified in Sec. \ref{2}, replacing the vacuum tagged-photon cut with Eq.~\eqref{bg_pro} gives the modified squared amplitude:
\begin{equation}
\left[1+n_\gamma(k)\right]
\overline{|\mathcal{M}_{\mathrm{ref}}|^2}.
\end{equation}
The observed photon corresponds to a definite occupation number, irrespective of which contribution to the coherent amplitude emits it.

We retain the photon coordinates $X=(y_\gamma,k_\perp)$ and $dX=dy_\gamma dk_\perp $ introduced in Sec.~\ref{2}. At fixed $X$, $n_\gamma(X)=n_\gamma(k)$ is independent of the remaining integration variables and can therefore be taken outside the integral. Denoting full integrand of the Eq.~\eqref{ref_diff} by $W(X,\Phi)$, we obtain
\begin{equation}
\frac{d\sigma_{\mathrm{ref}}}{dX}
=
\frac{d\sigma_{\mathrm{ref}}}
{dy_\gamma\,dk_\perp}
=
2\pi k_\perp
\int d\Phi\,W(X,\Phi),
\end{equation}
\begin{equation}
\frac{d\sigma_{\mathrm{bg}}}{dX}
=
2\pi k_\perp
\int d\Phi\,W(X,\Phi)
\left[1+n_\gamma(X)\right].
\end{equation}
The integrated reference samples a range of incident photon momenta and lepton configurations. At fixed $X$, however, every contribution carries the same final-state photon occupation number, preserving locality. Although the incident photon momentum fractions vary over the lepton phase space, this variation changes only the calculable reference weight; it does not integrate over different target-photon coordinates. Consequently,
\begin{equation}
R(X)
\equiv
\frac{
d\sigma_{\mathrm{bg}}/dX
}{
d\sigma_{\mathrm{ref}}/dX
}
=
1+n_\gamma(X),
\qquad
n_\gamma(X)=R(X)-1.
\end{equation}
This relation holds wherever the reference cross section is nonzero. No global parametrization of $n_\gamma(X)$ is required. Each resolvable photon momentum provides a local comparison between the observed spectrum and the theoretical reference.

The reference spectrum may itself exhibit pronounced structures arising from the nuclear photon luminosity and lepton-radiation kinematics. Within the adopted model, the cross section ratio removes these known process-dependent weights. A peak in the raw photon yield therefore does not, by itself, establish a corresponding peak in the occupation number. The relevant information is the additional contribution relative to the reference at that momentum. The extraction thus measures how strongly the surrounding photon ensemble enhances a specified radiative process. The extracted quantity is the occupation number defined in Sec.~\ref{2}. It can be related to the spatially integrated nuclear photon distribution through Eq.~\eqref{match}.

This result also permits tests at the same photon momentum using different lepton selections. Such selections change the event rate and reference weight, but, within the factorized model, each observed spectrum divided by its matched reference should yield the same occupation number. This comparison tests the assumed separation between radiation dynamics and the scalar statistical factor. Discrepancies exceeding the corresponding uncertainties would require further examination of the shared model assumptions or measurement treatment.

\section{Measurement Definition and Extraction Relation}

To verify that the pointwise inversion is not merely an idealized theoretical limit, we further test its practical robustness against potential ratio disruptions caused by realistic experimental complexities. At fixed observed photon variables $X$, experimental selections enter through a common kinematic cut function $C(X,\Phi)$ and detection efficiency $A(X,\Phi)$, with the remaining integration variables again denoted by $\Phi$. We define the reference kernel including cuts and efficiency as
\begin{equation}
\begin{aligned}
H_{C,A}(X,\Phi)
={}&
\frac{\alpha^3}{4\pi^2}
\frac{H_{\mathrm{BW}}}{\hat{s}^{\,2}}
\left[
x_1 f_1^{\gamma/A}
(x_1,k_{1\perp}^2)
\right]
\left[
x_2 f_1^{\gamma/B}
(x_2,k_{2\perp}^2)
\right]
\\
&\times C(X,\Phi)\,A(X,\Phi).
\end{aligned}
\end{equation}
For a measurement without photon momentum migration, let
\begin{equation}
W_{C,A}(X)
=
\int d\Phi\,
H_{C,A}(X,\Phi)\,
\delta^{(2)}
\!\left(
\boldsymbol{q}_\perp
-\boldsymbol{k}_{1\perp}
-\boldsymbol{k}_{2\perp}
\right).
\end{equation}
If the tagged-photon occupation factor remains multiplicative, the two spectra within the acceptance satisfy
\begin{equation}
\frac{d\sigma_{\mathrm{ref}}^{C,A}}{dX}
=
2\pi k_\perp W_{C,A}(X),
\end{equation}
\begin{equation}
\frac{d\sigma_{\mathrm{bg}}^{C,A}}{dX}
=
2\pi k_\perp
\left[1+n_\gamma(X)\right]
W_{C,A}(X).
\end{equation}
Defining their ratio as $R_{C,A}(X)$, we again obtain, wherever the reference cross section is nonzero,
\begin{equation}
R_{C,A}(X)-1=n_\gamma(X).
\end{equation}
Cuts and efficiencies may vary strongly with the lepton variables. Provided that both spectra receive the same treatment, however, the momentum coordinate of the extracted occupation remains fixed. If the experimental data have already been corrected for efficiency, the theoretical reference must match the corrected definition to avoid applying the efficiency twice. Selections may include lepton rapidity intervals, lepton transverse-momentum thresholds, and photon--lepton isolation requirements. These change the reference weight and normalization, but all selected events at fixed photon momentum retain the same occupation factor.

Using a bare-lepton definition, we describe additional untagged QED radiation by an azimuth-independent leading-logarithmic Sudakov exponent \(S_{\mathrm{LL}}(r_\perp;M_{ee},P_\perp)\) \cite{PhysRevD.108.116015,Klein_2019,Klein_2020,Li_2019,PhysRevD.101.034015}. In the small-$k_\perp$ forward-tagged-photon and small-recoil region specified above, the transverse momentum-conservation factor $\delta^{(2)}\!\left(\boldsymbol q_\perp-\boldsymbol k_{1\perp}-\boldsymbol k_{2\perp}\right)$ in the reference is replaced by \cite{Klein_2020,PhysRevD.101.034015,COLLINS1985199}
\begin{equation}
F(S_{\mathrm{LL}})\equiv
\int\frac{d^2\boldsymbol r_\perp}{(2\pi)^2}
e^{\,i\boldsymbol r_\perp\cdot
(\boldsymbol q_\perp-\boldsymbol k_{1\perp}-\boldsymbol k_{2\perp})}
e^{-S_{\mathrm{LL}}(r_\perp;M_{ee},P_\perp)}.
\end{equation}
Here, \(\boldsymbol r_\perp\) is Fourier conjugate to the transverse recoil. Within the assumed statistical factorization, the same radiative kernel enters both spectra and retains the multiplicative occupation factor at fixed \(X\). This replacement defines the radiative reference kernel at leading-logarithmic accuracy. The exponent applies in the small-recoil resummation region; outside that region, the result should be matched to a fixed-order calculation.

The Sudakov factor may depend on the lepton kinematics. Pointwise inversion requires that the corrected background spectrum retain the multiplicative form
\begin{equation}
\begin{aligned}
\frac{d\sigma_{\mathrm{bg,Sud}}^{C,A}}{dX}
&=
\left[1+n_\gamma(X)\right]
\frac{d\sigma_{\mathrm{ref,Sud}}^{C,A}}{dX}
\\
&=
\left[1+n_\gamma(X)\right]
2\pi k_\perp\int d\Phi\,
H_{C,A}(X,\Phi)\,
F(S_{\mathrm{LL}}).
\end{aligned}
\end{equation}

Additional coherent Coulomb rescattering of the produced leptons on the nuclei is neglected within the leading-power correlation approximation. This treatment is motivated by the eikonal cancellation for a compact charge-neutral lepton pair \cite{Klein_2019,Klein_2020} and, in the soft-tagged-photon limit, by soft-photon factorization onto the underlying dilepton amplitude \cite{PhysRev.110.974}. Corrections associated with the finite transverse size of the pair are beyond the present treatment.

A finite photon bin $B$ integrates over different occupation-number coordinates. Applying the same bin to numerator and denominator gives
\begin{equation}
\begin{aligned}
    R_{B,\mathrm{Sud}}^{C,A}
\equiv
1+n_{B,\mathrm{Sud}}^{C,A}&=\frac{
\displaystyle
\int_B dX\,
\frac{d\sigma_{\mathrm{bg,Sud}}^{C,A}}{dX}
}{
\displaystyle
\int_B dX\,
\frac{d\sigma_{\mathrm{ref,Sud}}^{C,A}}{dX}
}
\end{aligned}
\end{equation}
Thus, $n_{B,\mathrm{Sud}}^{C,A}$ is the occupation number averaged with the reference cross section as the weight. Identifying this average with the value at the bin center requires the occupation number to vary sufficiently slowly over the region carrying appreciable reference weight. Narrower bins improve momentum-space locality at the cost of event statistics. When the reference weight varies rapidly, the effective sampling point may also shift away from the geometric bin center. Reporting the reference weight or its moments helps quantify this shift.

Photon momentum migration introduces further averaging. Let $T(X_{\mathrm{rec}}\mid X)$ denote the detector momentum response averaged over the unobserved variables using the Sudakov-corrected reference kernel, where $X_{\mathrm{rec}}$ denotes the reconstructed photon variables. The reconstructed ratio, including all effects above, then satisfies
\begin{equation}
R_{\mathrm{rec}}(X_{\mathrm{rec}})-1
=
\frac{
\displaystyle
\int dX\,
T(X_{\mathrm{rec}}\mid X)\,
2\pi k_\perp
\int d\Phi\,
H_{C,A}(X,\Phi)\,
F(S_{\mathrm{LL}})\,
n_\gamma(X)
}{
\displaystyle
\int dX\,
T(X_{\mathrm{rec}}\mid X)\,
2\pi k_\perp
\int d\Phi\,
H_{C,A}(X,\Phi)\,
F(S_{\mathrm{LL}})
}.
\end{equation}
Even with a matched response, the extracted quantity may therefore average occupation numbers at different true photon momenta. An approximately pointwise interpretation requires a response sufficiently localized on the relevant momentum scales or a correction for migration. The response's width characterizes the momentum resolution of the extraction and should be assessed separately from the uncertainty in the estimated occupation number.

Combining these effects, the theoretical reference cross section in a reconstructed photon bin $B_{\mathrm{rec}}$ is
\begin{equation}
\begin{aligned}
\sigma_{\mathrm{ref}}^{(B_{\mathrm{rec}})}
={}&
\int_{B_{\mathrm{rec}}} dX_{\mathrm{rec}}
\int dX\,
T(X_{\mathrm{rec}}\mid X)\,
2\pi k_\perp
\\
&\times
\int d\Phi\,
H_{C,A}(X,\Phi)\,
F(S_{\mathrm{LL}}).
\end{aligned}
\end{equation}
Experimentally, the electron--positron pairs can be divided into nonoverlapping kinematic regions within the domain of our approximations, and the occupation number can be extracted independently at the same true photon momentum. Agreement among these extractions provides an internal test of the statistical factorization. Once differences in effective weights caused by finite binning and migration have been accounted for, compatible results can be combined with correlated uncertainties to improve precision by using events from all regions. Shared reference-spectrum systematic uncertainties must be retained, while the momentum resolution of the extraction should be evaluated independently.

\section{Summary and Outlook}

We propose extracting background photon occupation numbers using electron--positron pairs produced in situ in ultraperipheral heavy-ion collisions. A matched theoretical reference is constructed from the complete tree-level radiative Breit--Wheeler amplitude and the incident nuclear photon luminosity. Within the adopted scalar statistical factorization, the occupation number of the selected photon mode enters the cross section through a common Bose-enhancement factor. At fixed photon momentum, the cross section ratio therefore yields $n_\gamma=R-1$, without an assumed global parametrization of the target distribution.

Matched kinematic selections, acceptance treatment, and radiative corrections that preserve the separation of the statistical factor enter the common reference weight. Finite bins and detector momentum migration instead produce weighted averages over occupation numbers at different true momenta, with the sampled range determining the locality of the extraction.

Future work can develop phase-space or Wigner descriptions incorporating the space-time structure of the photon source \cite{PhysRevD.104.056011,SHI2025139317}, characterize how the radiation process samples a finite background, and test the domain of validity of scalar statistical factorization. Further assessments of momentum resolution, event rates, and theoretical reference uncertainties, using experimentally feasible photon and lepton-pair selections, will establish the local momentum scales that can be resolved in photon distributions.

\appendix
\section{Box Normalization, Momentum Independence, and Absolute Calibration with Two-Photon Processes}
\label{app1}

We first describe finite-box normalization and then explain how the $\gamma\gamma\to e^+e^-$ channel can determine the overall normalization when matching to nuclear photon distributions is valid.

In a fixed reference frame, consider a periodic box with volume $V$. The momentum-space cell associated with each mode is $\Delta^3 k=\frac{(2\pi)^3}{V},$ so that
\begin{equation}
\sum_{\boldsymbol{k}}
\longrightarrow
\frac{V}{(2\pi)^3}
\int d^3\boldsymbol{k}.
\end{equation}
All momentum modes are defined in the same periodic box; hence, $V$ is independent of the mode momentum. The photon number is
\begin{equation}
dN_\gamma
=
2\frac{V}{(2\pi)^3}
n_\gamma(k)\,d^3\boldsymbol{k}
=
f_\gamma(x,k_\perp)\,
dx\,d^2\boldsymbol{k}_\perp.
\end{equation}
Here, $n_\gamma(k)$ is the mode occupation number, and $f_\gamma(x,k_\perp)$ is the photon number distribution. Thus,
\begin{equation}
\label{eq:appA-mode-matching}
f_\gamma(x,k_\perp)
=
\frac{2V}{(2\pi)^3}
\frac{E_\gamma}{x}
n_\gamma(k)
=
V\widetilde{f}(x,k_\perp).
\end{equation}
The overall matching coefficient V can be estimated within a model, with an associated uncertainty, or calibrated directly from experimental data.

Consider a nuclear collision channel dominated by the two-photon subprocess $\gamma+\gamma\to e^++e^-.$ Its cross section in a fiducial region $\Omega$ is
\begin{equation}
\sigma_\Omega[f_A,f_B]
=
\int d\xi_A\,d\xi_B\,
K_\Omega(\xi_A,\xi_B)\,
f_A(\xi_A)f_B(\xi_B),
\end{equation}
where $d\xi_i=dx_i\,d^2\boldsymbol{k}_{i\perp}$ and $K_\Omega$ includes the hard cross section, final-state phase space, and the corresponding fiducial selections.

Using the occupation numbers extracted in the main text, we construct $\widetilde{f}$ according to Eq.~\eqref{eq:appA-mode-matching} and define the normalization integral
\begin{equation}
\widetilde{\sigma}_\Omega
[\widetilde{f}_A,\widetilde{f}_B]
=
\int d\xi_A\,d\xi_B\,
K_\Omega(\xi_A,\xi_B)\,
\widetilde{f}_A(\xi_A)
\widetilde{f}_B(\xi_B).
\end{equation}
For identical nuclei sharing a common matching constant, the ratio gives
\begin{equation}
\frac{
\sigma_\Omega[f_A,f_B]
}{
\widetilde{\sigma}_\Omega
[\widetilde{f}_A,\widetilde{f}_B]
}
=
V_{\mathrm{fit}}^2.
\label{matching}
\end{equation}
Here, the numerator is the measured absolute cross section, while the denominator is calculated from the extracted $\widetilde{f}$ distributions using the same kernel.

A simpler alternative is to compare the EPA $f_\gamma^{\mathrm{EPA}}(k)$ directly with $\widetilde{f}$. Under the common-constant matching assumption, define
\begin{equation}
V_{\mathrm{EPA}}(k)
\equiv
\frac{
f_\gamma^{\mathrm{EPA}}(k)
}{
\widetilde{f}(k)
}
=
\frac{(2\pi)^3x}{2E_\gamma}
\frac{
f_\gamma^{\mathrm{EPA}}(k)
}{
n_\gamma(k)}.
\end{equation}
If the EPA distribution coincides with the true photon number distribution, Eq.~\eqref{eq:appA-mode-matching} implies $V_{\mathrm{EPA}}(\xi)=V$, independent of momentum. The overall normalization can therefore be obtained by fitting this ratio to a constant over a kinematic region in which EPA is reliable. After calibration by either method, the same fitted constant reconstructs the full photon number distribution:
\begin{equation}
f_\gamma^{\mathrm{rec}}(k)
=
V_{\mathrm{fit}}\widetilde{f}(k).
\end{equation}
The reconstructed distribution can retain local structures beyond those represented by the EPA reference.

The two calibration methods serve complementary purposes. Calibration with a measured absolute two-photon cross section reduces dependence on the EPA shape and normalization, but requires a reliable cross section measurement, a corresponding theoretical kernel, and sufficient momentum coverage for the calibration integral. Direct calibration through the ratio of the EPA distribution to $\widetilde{f}$ is simpler and displays local shape differences relative to EPA, but its overall normalization depends on the EPA model and the fitting region. When sufficient experimental information is available, absolute-cross section calibration can therefore be used as the primary method, with the EPA ratio as a supplementary check. Otherwise, the latter provides a normalization estimate with an explicit model uncertainty.

\section{Beam-Collinear Hard Process and Polarization Averaging}
\label{app2}

This section presents the hard factor and the eight polarization contractions, with photon momenta along the beam axis. It explains why the cross section ratio \(R\) is independent of the photon polarization distributions under the stated angular integrations and gives the rationale for the kinematic selections.

We evaluate the massless hard tensor with the incident photons along opposite beam directions, the emitted photon parallel to \(k_1\), and the central leptons back to back in the transverse plane. Define the relative lepton transverse momentum and the total transverse recoil of the visible three-body system as $P_\perp=\frac{ p_{1\perp}- p_{2\perp}}{2},$ $ q_\perp= p_{1\perp}+ p_{2\perp}+k_\perp$. This leading-power hard approximation applies for \(k_{1\perp},k_{2\perp},k_\perp,q_\perp\ll P_\perp\) and \(k_\perp/k^+\ll1\). Physical transverse momenta are retained in the photon distributions and phase-space convolution.

After summing over the final-state lepton spins, define the hard tensor with open polarization indices as \(H_{abij\mu\nu}\). The index pairs \((a,b)\), \((i,j)\), and \((\mu,\nu)\) correspond to incident photon 1, incident photon 2, and the final-state photon, respectively. In the two-dimensional physical transverse-polarization basis, define \cite{Mulders_2001,Pisano_2013}
\begin{equation}
    U^{ab}=\delta_\perp^{ab},
\qquad
L_r^{ab}=2t_r^at_r^b-\delta_\perp^{ab},
\qquad
\boldsymbol t_r=(\cos\phi_r,\sin\phi_r).
\end{equation}
Here, \(r\) labels the three photon lines, and \(\phi_r\) is the corresponding polarization reference azimuth. The three letters in each hard-factor subscript specify the \(U\) or \(L\) contraction for these photon lines in the same order:
\begin{equation}
  H_{XYZ}
=
H_{abij\mu\nu}\,
X_1^{ab}Y_2^{ij}Z_3^{\mu\nu},
\qquad
X,Y,Z\in\{U,L\}.  
\end{equation}
Omitting the common factor $e^6$, the eight contractions are
\begin{equation}
H_{UUU}
=
H_{abij\mu\nu}
U^{ab}U^{ij}U^{\mu\nu}
=
\mathcal{N}\frac{F^2}{z^2},
\end{equation}
\begin{equation}
H_{UUL}
=
H_{abij\mu\nu}
U^{ab}U^{ij}L_3^{\mu\nu}
=
\mathcal{N}\frac{(1-z)^2}{z^2}
\cos(2\alpha_3),
\end{equation}
\begin{equation}
H_{ULU}
=
H_{abij\mu\nu}
U^{ab}L_2^{ij}U^{\mu\nu}
=
-\frac{16}{P_\perp^2}
\cos(2\alpha_2),
\end{equation}
\begin{equation}
H_{ULL}
=
H_{abij\mu\nu}
U^{ab}L_2^{ij}L_3^{\mu\nu}
=
-\frac{16(1-z)^2}{P_\perp^2}
\cos\!\left[2(\alpha_2+\alpha_3)\right],
\end{equation}
\begin{equation}
H_{LUU}
=
H_{abij\mu\nu}
L_1^{ab}U^{ij}U^{\mu\nu}
=
\mathcal{N}(1-z)^2
\cos(2\alpha_1),
\end{equation}
\begin{equation}
H_{LUL}
=
H_{abij\mu\nu}
L_1^{ab}U^{ij}L_3^{\mu\nu}
=
\mathcal{N}
\cos\!\left[2(\alpha_1-\alpha_3)\right],
\end{equation}
\begin{equation}
H_{LLU}
=
H_{abij\mu\nu}
L_1^{ab}L_2^{ij}U^{\mu\nu}
=
-\frac{16(1-z)^2}{P_\perp^2z^2}
\cos\!\left[2(\alpha_1+\alpha_2)\right],
\end{equation}
\begin{equation}
\begin{aligned}
H_{LLL}
&=
H_{abij\mu\nu}
L_1^{ab}L_2^{ij}L_3^{\mu\nu}
\\
&=
-\frac{8}{P_\perp^2z^2}
\Bigl\{
(1-z)^4
\cos\!\left[2(\alpha_1+\alpha_2+\alpha_3)\right]
\\
&\hspace{3.5em}
+
\cos\!\left[2(\alpha_1+\alpha_2-\alpha_3)\right]
+
z^4
\cos\!\left[2(\alpha_1-\alpha_2-\alpha_3)\right]
\Bigr\}.
\end{aligned}
\end{equation}
where \(x'=p_1^+/(p_1^++p_2^+)\) is the lepton momentum-sharing fraction, \(z=k^+/k_1^+\) is the emitted-photon longitudinal momentum fraction, and \(\theta\) is the azimuth of \(\boldsymbol P_\perp\), with \(0<x',z<1\). We use the abbreviations $\alpha_r=\phi_r-\theta,$ $F=z^2-z+1,$ $\mathcal N=\frac{8\left[x'^2+(1-x')^2\right]}{P_\perp^2x'(1-x')}.$

After integrating over all azimuths, only $H_{UUU}$ survives. Thus, within the selected kinematic region and with these angles unobserved, only the unpolarized photon distributions are required.

The near-axis, small-transverse-momentum selection probes the momentum structure of coherent nuclear photons. Photon distributions in this region are important for UPC and photon-TMD studies \cite{Li_2019,PhysRevD.101.034015,SHI2025139317}, and their local extraction also has methodological connections to studies of small-\(x\) gluon TMDs \cite{Metz_2011}. Selecting a finite near-axis region additionally aims to exploit the higher mode occupations expected there. For a given reference weight, integrated luminosity, and detection efficiency, a larger occupation number increases the selected event yield and improves the relative statistical precision of the extraction.

In the same reference frame, we require \(|y_1|<1\) and \(|y_2|<1\) and impose nonzero lepton-transverse-momentum and photon-energy thresholds. For a tagged photon with \(y_\gamma>3\), its opening angle with either central massless lepton is approximately \(35^\circ\) or larger, avoiding photon emission collinear to a lepton. Together with the thresholds, these selections keep the radiative propagator denominators away from zero and suppress the collinear enhancement of vacuum radiation and its associated angular sensitivity \cite{Klein_2020,DITTMAIER200069}. This selection also aims to reduce dilution of the enhancement signal by spontaneous emission outside the background field. Far from the nuclei, the enhancement factor approaches unity, so such contributions would reduce the occupation number inferred from \(R-1\). Residual outside-field contributions should therefore be assessed as a source of systematic dilution. Quantifying this effect requires an analysis of the space-time overlap between the radiation formation region and the background field. This geometric configuration may also inform future QCD studies of gluon radiation from quarks. Sufficient angular separation between the quark and gluon directions can facilitate the reconstruction of distinct jets.

\bibliography{ref}

\end{document}